\documentstyle[manuscript,aps]{revtex}

\begin{document}
\title{Kinetic temperature gradient driven modes in inhomogeneous plasmas}
\author{A. I. Smolyakov}
\address{Department of Fusion Research, Japan Atomic Energy Research Institute, \\
Naka, Japan\\
and Department of Physics and Engineering Physics, University of\\
Saskatchewan, Saskatoon, S7N5E2 Canada}
\author{M. Yagi}
\address{Research Institute for Applied Mechanics, Kyushu University, Japan}
\author{Y. Kishimoto}
\address{Department of Fusion Research, Japan Atomic Energy Research Institute, Naka,%
\\
Japan}
\date{\today }
\maketitle

\begin{abstract}
New unstable temperature gradient driven modes in an inhomogeneous plasma
are identified. These modes represent transient $\omega \simeq k_{\Vert
}v_{th}^{(e,i)}$ sound oscillations in magnetized plasma that are
kinetically destabilized via Landau interactions. Electron and ion sound
branches are unstable for large values of the Larmor radius parameter $%
k_{\bot }\rho _{e,i}\gg 1,$ respectively. The instability occurs due to a
specific plasma response that significantly deviates from Boltzmann
distribution in the region $k_{\bot }\rho _{i,e}\gg 1$ . Pacs: 52.35 Kt,
52.35 Qz
\end{abstract}

Small scale instabilities driven by temperature gradients are believed to be
responsible for particle and energy transport in a tokamak. The ion
temperature gradient driven turbulence has been identified \cite{dimits} as
a primary cause of ion thermal transport. The electron temperature gradient
driven modes are believed to be largely responsible for the electron
transport \cite{diii}. Both types of modes, the ion temperature gradient
(ITG) and electron temperature gradient (ETG), have been extensively studied
over last decades \cite
{tserkovnikov,rudakov,kadomtsev,galeev,mikhailovskii,jenko,kim} In basic
slab geometry these modes are essentially sound waves destabilized by
coupling to pressure fluctuations \cite
{tserkovnikov,rudakov,kadomtsev,galeev,mikhailovskii}. In toroidal geometry
the modes are driven by the unfavourable magnetic curvature rather than by
acoustic oscillations, though the mode may remain essentially slab-like in
the region of the weak/negative shear \cite{idomura,kishimoto}. In this work
we report on a new mechanism of the destabilization of the temperature
gradient driven modes. The mechanism is essentially related to the
wave-particle Landau interaction. However, the most important element is a
newly found regime of plasma response for large values of the Larmor radius
parameter, $k_{\bot }^{2}\rho _{\alpha }^{2}\gg 1$, $\alpha =(e,i).$ In this
regime plasma response in essential way deviates from usually assumed
Boltzmann distribution for $k_{\bot }^{2}\rho _{\alpha }^{2}\gg 1.$ As is
shown below modified plasma response occurs for transient (acoustic type)
modes whose eigen-frequency $\omega \simeq k_{\Vert }v_{th\alpha }$ does not
grow with increase of $k_{\bot }^{2}\rho _{\alpha }^{2}$, but rather remains
constant. As a result two new branches of the temperature gradient driven
modes occur for $k_{\bot }^{2}\rho _{i}^{2}\gg 1$ and $k_{\bot }^{2}\rho
_{e}^{2}\gg 1$. We call them ion and electron kinetic temperature gradient
(KTG) modes, respectively.

We consider shearless dispersion equation in a local approximation. Within
the local theory the parallel velocity and density perturbations for each
species are given by standard expressions (see e.g. \cite{lee,chenparker}) 
\begin{equation}
\frac{\widetilde{v}_{\alpha \Vert }}{v_{th\alpha }}=-\frac{e_{\alpha }%
\widehat{\phi }}{T_{\alpha }}s_{\alpha }D_{\alpha },  \label{vp}
\end{equation}

\begin{equation}
\frac{n_{\alpha }}{n_{0}}=-\frac{e_{\alpha }\phi }{T_{\alpha }}l_{\alpha }-%
\frac{e_{\alpha }\widehat{\phi }}{T_{\alpha }}D_{\alpha }.  \label{n}
\end{equation}
Here 
\begin{eqnarray}
D_{\alpha } &=&\left( 1-\frac{\omega _{n\alpha }}{\omega }\right) \left(
1+s_{\alpha }Z(s_{\alpha })\right) \Gamma _{0}(b_{\alpha })+\frac{\omega
_{T\alpha }}{\omega }s_{\alpha }\left( \frac{1}{2}Z(s_{\alpha })-s_{\alpha
}-s_{\alpha }^{2}Z(s_{\alpha })\right) \Gamma _{0}(b_{\alpha })  \nonumber \\
&&+\frac{\omega _{T\alpha }}{\omega }\left( 1+s_{\alpha }Z(s_{\alpha
})\right) \left( \Gamma _{0}(b_{\alpha })-\Gamma _{1}(b_{\alpha })\right)
b_{\alpha },  \label{d}
\end{eqnarray}
and 
\begin{equation}
l_{\alpha }=1-\left( 1-\frac{\omega _{n\alpha }}{\omega }\right) \Gamma
_{0}(b_{\alpha })-\frac{\omega _{T\alpha }}{\omega }\left( \Gamma
_{0}(b_{\alpha })-\Gamma _{1}(b_{\alpha })\right) b_{\alpha }.  \label{l}
\end{equation}
Various plasma parameters are defined as follows: $\omega _{n\alpha
}=-k_{y}cT_{\alpha }/e_{\alpha }B_{0}L_{n},$ $\omega _{T\alpha
}=-k_{y}cT_{\alpha }/e_{\alpha }B_{0}L_{T\alpha
},L_{n}^{-1}=-n_{0}^{-1}\partial n_{0}/\partial x$, $L_{T\alpha }=-T_{\alpha
}^{-1}\partial T_{\alpha }/\partial x,$ $s_{\alpha }=\omega /k_{\Vert
}v_{th\alpha },$ $b_{\alpha }=k_{\bot }^{2}\rho _{\alpha }^{2}/2,$ $%
v_{tha}^{2}=2T_{\alpha }/m_{\alpha }$, $\rho _{\alpha }=v_{tha}m_{\alpha
}c/(e_{\alpha }B_{0});$ $\Gamma _{0,1}(b)=I_{0,1}\exp (-b)$, and $Z(s)$ is
the standard plasma dispersion function. We have introduced auxiliary
potential $\widehat{\phi }=\phi -\omega /(k_{\Vert }c)A$, where $\phi $ is
the electrostatic potential and $A$ is the magnetic vector potential.

The vector potential $A$ can be found from the Ampere's law

\begin{equation}
J_{\Vert }=en(v_{i\Vert }-v_{e\Vert })=-\frac{c}{4\pi }\nabla _{\bot }^{2}A.
\label{a}
\end{equation}
By using (\ref{vp}) and (\ref{a}) one obtains

\begin{equation}
A=-\frac{2s_{e}^{2}\left( D_{e}\tau +D_{i}\right) }{k_{\bot }^{2}\delta
^{2}-2s_{e}^{2}\left( D_{e}\tau +D_{i}\right) }\frac{k_{\Vert }c}{\omega }%
\phi ,
\end{equation}
and 
\begin{equation}
\widehat{\phi }=\phi \frac{k_{\bot }^{2}c^{2}/\omega _{pe}^{2}}{k_{\bot
}^{2}\delta ^{2}-2s_{e}^{2}\left( D_{e}\tau +D_{i}\right) },  \label{phib}
\end{equation}
for the parallel potential $\widehat{\phi }.$ Here $\tau =T_{e}/T_{i}$, and $%
\delta ^{2}=c^{2}/\omega _{pe}^{2}.$

By using the Poisson equation, 
\begin{equation}
-\nabla _{\bot }^{2}\phi =4\pi e(n_{i}-n_{e}),
\end{equation}
and expressions (\ref{n}) and (\ref{phib}) we obtain general local
dispersion relation 
\begin{eqnarray}
&&k_{\bot }^{2}\delta ^{2}\left( l_{i}\tau +l_{e}+D_{i}\tau +D_{e}\right)
-2s_{e}^{2}\left( D_{i}\tau +D_{e}\right) \left( l_{i}\tau +l_{e}\right) 
\nonumber \\
&=&-k_{\bot }^{2}\lambda _{D}^{2}\left( k_{\bot }^{2}\delta
^{2}-2s_{e}^{2}\left( D_{i}\tau +D_{e}\right) \right) ,  \label{deq}
\end{eqnarray}
where $\lambda _{D}^{2}$ is the Debye lenghth, $\lambda _{D}^{2}=T_{e}/(4\pi
n_{0}e^{2}).$ This general dispersion equation describes both ITG and ETG
modes as well as new kinetic temperature gradient driven modes.

To illustrate existence of a new unstable mode we plot a solution of the
dispersion equation (\ref{deq}) as a function of the normalized ion Larmor
radius parameter $\rho _{i}/L_{n}.$ The following parameters are fixed $%
k_{y}\rho _{i}=\sqrt{2}\times 0.3$, $k_{x}\rho _{i}=\sqrt{2}\times 0.1$, $%
\beta =2\times 10^{-4}$, $\rho _{i}/L_{Ti}=\sqrt{2}\times 0.1$, $\rho
_{i}/L_{Te}=\sqrt{2}\times 0.01$, $k_{\Vert }\rho _{i}=\sqrt{2}\times 0.002$%
, $\tau =1,$ We have chosen the above plasma parameters to be exactly the
same as in Ref. 17. The plasma pressure parameter $\beta $ is defined as $%
\beta =8\pi n_{0}T_{i}/B_{0}^{2}.$ [Note that our definition of $v_{th\alpha 
\text{ }}^{2}$and $\beta $ differ differ from those in Ref. 17 by a factor
of 2.] Figures 1a and 1b show the real $\omega _{r}$ and imaginary $\omega
_{i}$ parts of the eigen frequency, respectively. It is already evident from
Figs 1a and 1b that there are exist two distinct eigenmodes: in Fig. 1b the
left curve is the unstable ITG branch and the right curve is the unstable
drift mode branch \cite{chenparker} (in terminology of Ref. 17). Our results
in Figs 1a and 1b are in complete agreement with Ref. 17.

To clarify the nature of the unstable branch we solve the dispersion
equation as a function of the $k_{y}\rho _{i}$ parameter for fixed $\beta
=2\times 10^{-4}$, $\rho _{i}/L_{n}=\sqrt{2}\times 10^{-2}$, $\rho
_{i}/L_{Ti}=\sqrt{2}\times 10^{-1}$, $\rho _{i}/L_{Te}=\sqrt{2}\times 10^{-1}
$, $k_{x}\rho _{i}=\sqrt{2}\times 10^{-1}$, $\tau =1$. We also take $\lambda
_{D}=0$ for this case. The mode growth rate and mode frequency normalized to 
$k_{\Vert }v_{the}$ are shown in Figs. 2a and 2b, respectively. Two new
unstable branches exist in the regions $k_{y}\rho _{i}\geq 1$ and $k_{y}\rho
_{e}\geq 1$. The standard ITG mode has a peak growth rate around $k_{y}\rho
_{i}\simeq 1$, and the standard ETG mode has a growth rate peaked at $%
k_{y}\rho _{i}\simeq 40$ (corresponding to $k_{y}\rho _{e}\simeq 1)$. A new
ion mode (ion KTG) has a maximum growth rate around $k_{y}\rho _{i}\simeq 5$%
, and a new electron mode (electron KTG) saturates to a constant growth rate
quickly for large $k_{y}\rho _{i}>200$ ( $k_{y}\rho _{e}>5).$ It is
essential to note that new modes exist in the regions of large Larmor radius
parameters (respectively, $k_{y}\rho _{i}$ for the ion KTG, and $k_{y}\rho
_{e}$ for the electron ETG). As we show below, kinetic temperature gradient
modes occur due to a peculiar behaviour of the plasma response in the region
of large $k_{y}\rho _{\alpha }$. Let us consider this response in more
details.

A standard notion is that for large values of the Larmor radius parameter, $%
k_{y}\rho _{\alpha }>1$, the density response of the respective plasma
component is Boltzmann due to decaying asymptotics of $\Gamma
_{0,1}(b_{\alpha })\sim 1/\sqrt{b_{a}}$ for large $b_{\alpha }$, so that we
have from (\ref{d}-\ref{l}) $D_{\alpha }\ll 1,l_{\alpha }\simeq 1$. This, in
fact, implicitly assumes that the ratio of $\omega _{n\alpha }/\omega $ is
finite for large $k_{y}\rho _{\alpha }$. In turn, this requires that the
mode eigen-frequency increases with $k_{y}\rho _{\alpha }$ (linearly or
faster). The latter is definitely true for drift wave type modes, where $%
\omega \sim \omega _{*\alpha }$. However, temperature gradient driven modes,
are basically sound waves whose frequency is of the order of the transient
(or sound) frequency, $k_{\Vert }v_{th\alpha }$. If the parameter $k_{\Vert
}v_{th\alpha }$ remains constant (so $\omega $ is approximately constant),
the ratio $\omega _{n\alpha }/\omega $ will be increasing with $k_{y}\rho
_{\alpha }\,$that may compensate for the decaying $1/\sqrt{b_{a}}$ factor
from $\Gamma _{0,1}(b_{\alpha })$. Then, for large $k_{y}\rho _{\alpha }\gg 1
$ we may write in the leading order 
\begin{equation}
\frac{\omega _{n\alpha }}{\omega }\Gamma _{0}(b_{\alpha })=\frac{1}{2\sqrt{%
\pi }}\frac{e_{\alpha }}{e}\frac{v_{th\alpha }}{\omega L_{n}},
\end{equation}
and 
\begin{equation}
\frac{\omega _{T\alpha }}{\omega }\left( \Gamma _{0}(b_{\alpha })-\Gamma
_{1}(b_{\alpha })\right) b_{\alpha }=\frac{1}{4\sqrt{\pi }}\frac{e_{\alpha }%
}{e}\frac{v_{th\alpha }}{\omega L_{T\alpha }}.
\end{equation}
By using these asymptotics for the electron KTG mode in the region $%
k_{y}\rho _{e}\gg 1$ and $k_{y}\rho _{i}\gg 1$ we obtain the following
response functions

\begin{equation}
l_{i}=1-\frac{1}{2\sqrt{\pi }}\frac{v_{ti}}{\omega L_{n}}+\frac{1}{4\sqrt{%
\pi }}\frac{v_{ti}}{\omega L_{Ti}},  \label{lis}
\end{equation}
\begin{equation}
l_{e}=1+\frac{1}{2\sqrt{\pi }}\frac{v_{te}}{\omega L_{n}}-\frac{1}{4\sqrt{%
\pi }}\frac{v_{te}}{\omega L_{Te}},  \label{les}
\end{equation}
\begin{equation}
D_{i}=-\frac{1}{4\sqrt{\pi }}\frac{v_{ti}}{\omega L_{n}}\frac{1}{s_{i}^{2}}+%
\frac{1}{4\sqrt{\pi }}\frac{v_{ti}}{\omega L_{Ti}}\frac{1}{s_{i}^{2}},
\end{equation}
\begin{eqnarray}
D_{e} &=&-\frac{1}{2\sqrt{\pi }}\frac{v_{te}}{\omega L_{n}}\left(
1+s_{e}Z(s_{e})\right) +\frac{1}{2\sqrt{\pi }}\frac{v_{te}}{\omega L_{Te}}%
s_{e}\left( \frac{1}{2}Z(s_{e})-s_{e}-s_{e}^{2}Z(s_{e})\right)   \nonumber \\
&&+\frac{1}{4\sqrt{\pi }}\frac{v_{te}}{\omega L_{Te}}\left(
1+s_{e}Z(s_{e})\right) .  \label{des}
\end{eqnarray}
The electron KTG mode is a destabilized electron sound wave with $\omega
\simeq k_{\Vert }v_{the}$ so we retain a full plasma dispersion function in $%
D_{e}$ to account for the electron Landau interactions. In this region $%
\omega /k_{\Vert }v_{thi}\gg 1$ so we can simplify $Z(s_{i})$ functions.
Solution of dispersion equation (\ref{deq}) with simplified response
functions (\ref{lis}-\ref{des}) and $\lambda _{D}=0$ is shown in Fig. 2 by
squares. We note here, that in fact plasma response is not given exactly by
expressions (\ref{vp}-\ref{n}) but contains additional terms of the order of 
$\rho _{i}/L_{n}$ \cite{smolyakov}$.$ These additional terms do not affect
asymptotic expressions (\ref{lis}-\ref{des}) as long as $\rho _{i}/L_{n}<1.$

A similar expansion can be made in the region $k_{y}\rho _{i}\gg 1$. Then
the ion response for the ion KTG branch becomes 
\begin{equation}
l_{i}=1-\frac{1}{2\sqrt{\pi }}\frac{v_{ti}}{\omega L_{n}}+\frac{1}{4\sqrt{%
\pi }}\frac{v_{ti}}{\omega L_{Ti}},
\end{equation}

\begin{eqnarray}
D_{i} &=&\frac{1}{2\sqrt{\pi }}\frac{v_{ti}}{\omega L_{n}}\left(
1+s_{i}Z(s_{i})\right) -\frac{1}{4\sqrt{\pi }}\frac{v_{ti}}{\omega L_{Ti}}%
s_{i}\left( \frac{1}{2}Z(s_{i})-s_{i}-s_{i}^{2}Z(s_{i})\right)  \nonumber \\
&&-\frac{1}{2\sqrt{\pi }}\frac{v_{ti}}{\omega L_{Ti}}\left(
1+s_{i}Z(s_{i})\right) .
\end{eqnarray}

The electron response functions can be simplified in this region by
explicitly using $k_{\bot }\rho _{e}<1$ and $\omega /k_{\Vert }v_{te}<1$.
Then one obtains 
\begin{equation}
l_{e}=1-\left( 1-\frac{\omega _{ne}}{\omega }\right) ,\   \label{le2}
\end{equation}
\begin{equation}
D_{e}=\left( 1-\frac{\omega _{ne}}{\omega }\right) \left( 1-2s_{e}^{2}+is_{e}%
\sqrt{\pi }\right) 1+\frac{\omega _{Te}}{\omega }s_{e}\left( -2s_{e}+i\frac{%
\sqrt{\pi }}{2}\right) ,
\end{equation}

\begin{equation}
l_{e}=1-\left( 1-\frac{\omega _{ne}}{\omega }\right) \left( 1-b_{e}\right) -%
\frac{\omega _{Te}}{\omega }b_{e},
\end{equation}
\begin{eqnarray}
D_{e} &=&\left( 1-\frac{\omega _{ne}}{\omega }\right) \left(
1-2s_{e}^{2}+is_{e}\sqrt{\pi }\right) \left( 1-b_{e}\right) +\frac{\omega
_{Te}}{\omega }s_{e}\left( -2s_{e}+i\frac{\sqrt{\pi }}{2}\right) (1-b_{e}) 
\nonumber \\
&&+\frac{\omega _{Te}}{\omega }\left( 1-2s_{e}^{2}+is_{e}\sqrt{\pi }\right)
\ b_{e}.  \label{de2}
\end{eqnarray}

Mode growth rate and real frequency obtained with (\ref{le2}-\ref{de2}) and (%
\ref{deq}) with $\lambda _{D}=0$ are shown in Fig. 2 by circles.

Recently, it has been emphasised that the ETG modes can be significantly
modified by a finite value of the Debye length parameter \cite{idomura}.
Electron KTG mode is strongly stabilized in a high temperature plasma by the
finite Debye length as shown in Fig. 3. Note that effect of the Debye length
screening may be different for the negative shear plasma \cite{idomura}.

Though the modified ion response for large ion Larmor radius is criticial
for the ion KTG mode, the mode itself is destabilized by the electron
temperature gradient as demonstrated in Fig 4. Ion KTG mode exists for large
gradients of the electron temperarure (dashed line, $\eta _{e}=5).$ The mode
growth rate is significantly reduced for smaller values of $\eta _{e}$
(solid line, $\eta _{e}=0.5).$ Thus we may conclude that ion KTG mode is in
fact a hybrid mode, in which both ion and electron response are essential.
Electromagnetic effects are expected to be important for this mode existing
in the intermediate region between standard ITG and ETG modes. Though for
larger $\eta _{e}$ we observe a standard finite $\beta $ suppression (curve
labeled by diamonds in Fig. 5), for low $\eta _{e}=0.5$ , finite plasma
pressure effectively leads to the extension of the ITG instability into the
shorter wavelengths (curve labeled by squares in Fig. 5). It is interesting
to note that for $\beta =0.1$ taken for this case, the skin depth size $%
c/\omega _{pe}\simeq 0.1$ $\rho _{i}$, which makes $k_{y}c/\omega _{pe}\leq
1 $ in the region of the unstable ion KTG.

In summary, we have identified two new branches of the temperature gradient
driven modes. These modes are closely related to acoustic type modes that
are destabilized by the temperature gradient effects. A modified plasma
response in the region of $k_{y}\rho _{\alpha }\gg 1$ is essential for new
instability. There is certain similarity of the KTG modes to the universal
instability of drift waves that is driven by density gradient and
wave-particle interaction \cite{berk}.  The KTG modes considered in this
work will be affected by a finite value of the magnetic shear but noting a
special role of the temperature gradients and the modified plasma responce
for $k_{y}\rho _{\alpha }\gg 1$ it remains unclear whether these modes  can
be stabilized by the shear similarly to the universal instability \cite
{ross,tsang}.  One can expect that KTG modes can persist in the region of
weak and/or negative magnetic shear where the mode structure is essentially
that of a shearless slab \cite{kishimoto}. Toroidal effects will further
modify the KTG modes but are not expected to cause their stabilization \cite
{hirose}. Generally, investigation of a general case requires nonlocal
integral analysis \cite{dong,dong2,idomura} that is left for future work.

Authors would like to acknowledge useful discussions with Y. Idomura, J. Q.
Li, S.I. Itoh, and K. Itoh. We are also grateful to Dr. H. Ninomiya and Dr.
A. Kitsunezaki for their continuous encouragement. This research was
supported by Natural Sciences and Engineering Research Council of Canada, by
the Grant-in-Aid for Scientific Research and the Grant-in-Aid for
International Scientific Research of Ministry of Education, Culture, Sports,
Science and Technology of Japan and by collaboration program of Research
Institute for Applied Mechanics of Kyushu University.

\newpage

\centerline{FIGURES}

FIG.1. Normalized (to the ion cyclotron frequency) frequency (a) and growth
rate (b) of the ITG and unstable drift wave modes versus the normalized ion
Larmor radius $\rho _{i}/L_{n}.$ The solid line corresponds to the ITG
branch, while the dashed line is that of the destabilized drift wave branch.
Parameters are the same as those in Ref. $\beta =2\times 10^{-4}$, $\rho
_{i}/L_{Ti}=\sqrt{2}\times 10^{-1}$, $\rho _{i}/L_{Te}=\sqrt{2}\times 10^{-2}
$, $k_{x}\rho _{i}=\sqrt{2}\times 10^{-1}$, $k_{y}\rho _{i}=3\sqrt{2}\times
10^{-1}$, $k_{\Vert }\rho _{i}=2\sqrt{2}\times 10^{-3}.$

FIG.2. Normalized wave frequency (a) and growth rate (b) for the ITG , ETG
modes and kinetic modes. Plasma parameters: $\beta =2\times 10^{-4}$, $\rho
_{i}/L_{n}={\sqrt{2}}\times 10^{-2}$, $\rho _{i}/L_{Ti}={\sqrt{2}}\times
10^{-1}$, $\rho _{i}/L_{Te}={\sqrt{2}}\times 10^{-1}$, $k_{x}\rho _{i}={%
\sqrt{2}}\times 10^{-1}$, $k_{\Vert }\rho _{i}=2\sqrt{2}\times 10^{-3},$ $%
\tau =1$. Solid line -- standard ITG and ion KTG; dashed line -- standard
ETG and electron KTG.

FIG.3. Stabilization of the Kinetic Electron Temperature Gradient mode by
the Debay screening effect: solid line -- $\lambda _{D}/\rho _{e}=0,$
dotted-dashed line -- $\lambda _{D}/\rho _{e}=0.70$, dotted line -- $\lambda
_{D}/\rho _{e}=2.21.$ Plasma parameters: $\beta =2\times 10^{-4}$, $\rho
_{i}/L_{n}={\sqrt{2}}\times 10^{-2}$, $\rho _{i}/L_{Ti}={\sqrt{2}}\times
10^{-1}$, $\rho _{i}/L_{Te}={\sqrt{2}}\times 10^{-1}$, $k_{x}\rho _{i}={%
\sqrt{2}}\times 10^{-1}$, $k_{\Vert }\rho _{i}=2\sqrt{2}\times 10^{-3},$ $%
\tau =1$.

FIG.4. Effect of the electron temperature gradient on KTG modes. Solid line
-- $\rho _{i}/L_{Te}={\sqrt{2}}\times 10^{-2}$ ($\eta _{e}=0.5)$, dashed
line $\rho _{i}/L_{Te}={\sqrt{2}}\times 10^{-1}$ ($\eta _{e}=5)$. Plasma
parameters: $\beta =2\times 10^{-4}$, $\rho _{i}/L_{n}={\sqrt{2}}\times
10^{-2}$, $\rho _{i}/L_{Ti}={\sqrt{2}}\times 10^{-1}$, $k_{x}\rho _{i}={%
\sqrt{2}}\times 10^{-1}$, $k_{\Vert }\rho _{i}=2\sqrt{2}\times 10^{-3},$ $%
\tau =1$.

FIG.5. Effect of a finite plasma pressure on the KTG modes. $\ $ Triangles
-- $\rho _{i}/L_{Te}={\sqrt{2}}\times 10^{-1}$, $\beta =2\times 10^{-4}$;
diamonds squares -- $\rho _{i}/L_{Te}={\sqrt{2}}\times 10^{-1}$, $\beta
=0.1; $ circles $\rho _{i}/L_{Te}={\sqrt{2}}\times 10^{-2}$, $\beta =2\times
10^{-4}$, squares -- $\rho _{i}/L_{Te}={\sqrt{2}}\times 10^{-2}$, $\beta
=0.1.$ Other plasma parameters: -- $\rho _{i}/L_{n}={\sqrt{2}}\times 10^{-2}$%
, $\rho _{i}/L_{Ti}={\sqrt{2}}\times 10^{-1}$, $k_{x}\rho _{i}={\sqrt{2}}%
\times 10^{-1}$, $k_{\Vert }\rho _{i}=2\sqrt{2}\times 10^{-3},$ $\tau =1$.

\end{document}